\documentclass[11pt]{article}
\usepackage{epsfig}
\usepackage[utf8]{inputenc} 
\usepackage{url}

\begin{document}

\newcommand{\bm}[1]{\mbox{\boldmath$#1$}}

\def\mvec#1{{\bm{#1}}}   

\title{The Waves and the Sigmas\\
(To Say Nothing of the 750\,GeV Mirage)\footnote{Note
based on the invited talk 
{\em Claims of discoveries based on sigmas}
 at MaxEnt 2016 (Ghent, Belgium, 15 July 2016) 
and on seminars and courses to PhD students in the 
first half of 2016. 
}
}

\author{G.~D'Agostini \\
Universit\`a ``La Sapienza'' and INFN, Roma, Italia \\
{\small (giulio.dagostini@roma1.infn.it,
 \url{http://www.roma1.infn.it/~dagos})}
}

\date{}

\maketitle

\begin{abstract}
This paper shows how 
p-values do not only create, as well known, 
wrong expectations in the case of
flukes, but they might also dramatically diminish  
the `significance' of {\em most likely} genuine signals.
As real life examples, the 2015 first detections 
of gravitational waves are discussed.
 The March 2016 statement of the 
American Statistical Association, warning
scientists about interpretation and misuse of p-values,
is also reminded and commented. (The paper
is complemented with some remarks on past,
recent and future claims of discoveries {\em based 
on sigmas} from Particles Physics.)
\end{abstract}

\vspace{0.6cm}
\section{Introduction}
On February 11 the LIGO-Virgo collaboration 
announced the detection of Gravitational Waves (GW).
They were emitted about one billion years ago 
by a Binary Black Hole (BBH) 
merger and reached Earth on  September 14, 2015. 
The claim, as it appears in the 
`discovery paper'\,\cite{LIGO-P150914_Detection}
and stressed  in press releases and seminars,  was based on  
``$> 5.1\,\sigma$ significance.''  Ironically, shortly after, 
on March 7 the American Statistical
Association (ASA) came out (independently) with a strong statement warning
scientists about interpretation and misuse of p-values\,\cite{Asa_statement}.
As promptly reported by Nature\,\cite{Nature-ASA}, 
 ``this is the first time that the 177-year-old ASA 
has made explicit recommendations on such a 
foundational matter in statistics, says executive director Ron Wasserstein. 
The society's members had become increasingly concerned that the P value 
was being misapplied in ways that cast doubt on statistics generally, 
he adds.'' 

In June we have finally learned\,\cite{LIGO-June2016} 
that another `one {\em and a half'} gravitational waves from 
Binary Black Hole mergers were also observed in 2015, where
by the `half' I refer to the October 12 event, highly {\em believed} by
the collaboration to be a gravitational wave, although having
only 1.7\,$\sigma$ {\em significance} 
and therefore classified just 
as LVT (LIGO-Virgo Trigger) instead of GW. 
However, another figure of merit 
has been provided by the collaboration for each event, a number
based on probability theory and that tells how much we 
\underline{must} modify the relative beliefs
of two alternative hypotheses in the light of the 
experimental information. This number, at my knowledge
never even mentioned  
in press releases or seminars to large audiences, 
is the Bayes factor (BF), whose meaning is 
easily explained: if you considered {\em \`a priori} 
two alternative hypotheses equally likely, 
a BF of 100 changes your odds to 100 to 1;
if instead you considered one hypothesis rather unlikely,
let us say your odds were 1 to 100, a BF of $10^4$ turns
them the other way around, that is 100 to 1. 
You will be amazed to learn that even the ``1.7 sigma'' 
LVT151012 has a BF of the order of $\approx 10^{10}$, 
considered a very strong evidence in favor of the 
hypothesis  ``Binary Black Hole merger'' against the alternative
hypothesis ``Noise''. (Alan Turing 
would have called the evidence provided by such
an huge `Bayes factor,' or what I. J. Good would have preferred to call
``Bayes-Turing factor''\,\cite{Good-BTF},\footnote{Note that 
Eq. (1) in \cite{Good-BTF} clearly contains a typo, or it has got
a problem in the scanning of the document, since $P(E\,|\,H)/P(E\,|\,H)$
makes no sense in that equation and  it should have been 
 $P(E\,|\, H)/P(E\,|\, \overline H)$, 
where $H$ and $\overline H$ stand for `complementary' 
(formally ``exhaustive, mutually exclusive'') hypotheses.
The equation should then read
\begin{eqnarray*}
\frac{O(H\,|\, E)}{O(H)} &=& \frac{P(E\,|\, H)}{P(E\,|\, \overline H)}\,,
\end{eqnarray*}
where $O(H)$ and $O(H\,|\, E)$ are {\em prior} and {\em posterior}
odds, i.e., respectively, 
$O(H\,|\, E) = {P(H\,|\, E)}/{P(\overline H\,|\, E)}$ and 
$O(H) = {P(H)}/{P(\overline H)}$. 
Eq.\,(1) of \cite{Good-BTF} would then result into  
\begin{eqnarray*}
\frac{P(H\,|\, E)/P(\overline H\,|\, E)}{P(H)/P(\overline H)}
&=& \frac{P(E\,|\, H)}{P(E\,|\, \overline H)} 
\end{eqnarray*}
or 
\begin{eqnarray*}
\frac{P(H\,|\, E)}{P(\overline H\,|\, E)} &=&
 \frac{P(E\,|\, H)}{P(E\,|\, \overline H)} \cdot \frac{P(H)}{P(\overline H)}\,,
\end{eqnarray*}
in words
\begin{eqnarray*}
\mbox{\bf posterior odds} &=& \mbox{\bf Bayes factor} \times 
\mbox{\bf prior odds}
\end{eqnarray*}
(For log representation of odds and Bayes factors 
 see section 2
and appendix E of \cite{Columbo} and references therein, 
although at that time 
Turing's contributions, as well as `bans' and `decibans', 
were unknown to the author, who arrived at the same conclusion
of Turing's 1 deciban as rough estimate of {\em human resolution} 
to {\em judgement leaning} and {\em weight of evidence}
   -- table 1 in page 13  and text just below it.) 
}
100 {\em deciban}, well above the  17 deciban
threshold  considered by the team 
at Bletchley Park during World War II 
to be reasonably confident of having cracked the 
daily Enigma key\,\cite{McGrayne}.) 

In the past I have been writing quite a bit  on how 
`statistical' considerations based on p-values 
tend to create wrong expectations in 
frontier physics (see e.g. \cite{BR} and \cite{Badmath}). 
The main purpose of this paper is the opposite, i.e. to show how 
p-values might relegate to the role of a possible fluke
what is most likely a genuine finding. 
In particular, the solution of the apparent
paradox of how a marginal `1.7 sigma effect' 
could have a huge BF such as $10^{10}$ (and virtually even much more!)
is explained in a didactic way. 

\section{Preamble}
Since this paper can be seen as the sequel of Refs. \cite{Vulcano}
and \cite{Badmath}, with the basic considerations already
expounded in \cite{BR}, for the convenience of the reader I 
shortly summarize the main points maintained there.
\begin{itemize}
\item The ``essential problem of the experimental method'' 
      is nothing but solving ``a problem in the probability of causes'', 
      i.e.  ranking  in credibility the hypotheses
      that are considered to be possibly responsible of the observations,
      (quotes by Poincar\'e\,\cite{Poincare}).\,\footnote{Instead, 
      ``making statistics'', i.e. to 
      describe and summarize data, has never been the {\em primary} interest
      of physicists as well as of many other scientists, although it is
      certainly useful for a variety of reasons.}\\
      $[$There is indeed
       no conceptual difference between ``comparing hypotheses'' 
      or ``inferring the value'' of a physical quantity, the two problems
      only differing in the numerosity of hypotheses, {\em virtually} 
      infinite in the latter case, when the physical quantity is
      {\em assumed}, for mathematical 
      convenience,%
      \footnote{``{\em No mathematical squabbles}'' was John Skilling's 
      mantra in his recent tutorial  at MaxEnt 2016, 
      in which he was stressing the importance
      to restart thinking, at least ``initially'', in terms 
      of ``finite target'', ``finite partitioning'' and
      integers\,\cite{Skilling_ME2016}.} 
      to assume values with 
      continuity.$]$
\item The deep source of uncertainty in inference is due 
      to the fact that (apparently)  identical {\em causes} 
      might produce different effects, due to
      {\it internal} (intrinsic) probabilistic aspects of the theory, 
      as well as to {\em external} factors (think at measurement errors).
\item Humankind is used to live -- and survive --
      in conditions of uncertainty and therefore the human mind
      has developed a mental 
      `category' to handle it: {\em probability}, 
      meant as degree of belief. This is also valid when we `make science',
      since ``it is scientific only to say what is more likely 
      and what is less likely'' (Feynman\,\cite{Feynman_scientific}).
\item {\em Falsificationism} 
      can be recognized as an attempt to extend 
      the classical {\it proof by contradiction} of classical logic 
      to the experimental method, but it {\em simply fails} when 
      stochastic  (either internal or external) effects might occur. 
\item The further extension of falsificationism from  
      {\em impossible} effects to {\em improbable} effects is 
      simply deleterious.
\item The invention of p-values can be seen as 
      an attempt to overcome the evident problem occurring in the case 
      of a large number of effects ({\em virtually infinite} when 
      we make measurements): any observation has a very small probability
      in the light of whatever hypothesis is considered, and then 
      it `falsifies' it.   
\item Logically the previous extension (``observed effect'' $\rightarrow$
      ``all possible effects equally or less probable than the observed one'')
       does not hold water.
      (But it seems that for many practitioners logic is optional --
      the reason why ``p-values {\em often work}''\,\cite{BR} 
      will be discussed in 
      section \ref{sec:p-values_BF}.)
\item In practice p-values are routinely misinterpreted by most 
      practitioners and scientists, and 
      incorrect interpretations of the data are spread around over the 
      media\footnote{Sometimes 
      scientists say they reported 
      ``the right thing''
      (i.e. just the p-value), but it was journalist's fault 
      to misinterpret them. But, as I have documented in my writings, 
      often are the official statement of laboratories, of collaboration
      spokespersons, or of prominent physicists to confuse p-values
      with probabilities of hypotheses, as you can e.g. find in 
      \cite{Badmath} and, more extensively, in 
      \url{http://www.roma1.infn.it/~dagos/badmath/index.html#added}.
      A suggestion to laymen is that, ``instead of heeding 
      impressive-sounding statistics, we 
      should ask what scientists actually believe''\,\cite{PhilipBall}.
      } (for recent 
        examples, related to the LHC presumptive 
        750\,GeV di-photon signal (see
        e.g. \cite{NYT_Dec2015,Mirror_Dec2015,LeScienze_Dic2015,Bloomberg_Dec2015,ScienceAlert_Dec2015} and footnote \ref{fn:LHC_750GeV} for later comments.).
\item The reason of the misunderstandings is that 
      p-values (as well as other outcomes from other methods of 
      the dominating `standard statistics', including 
      {\em confidence intervals}\,\cite{BR}), 
      do not reply to the very question human minds
      {\em by nature} ask for, i.e. which hypothesis is more or less
      believable (or how likely the `true' value
      of a quantity lies within a given interval).  
      For this reason 
      I am afraid p-values (or perhaps a new invention by statisticians) 
      will still be misinterpreted 
      and misused despite the 2016 ASA statement, as I will argue at the 
      end of section \ref{ss:Principia}). 
\item Given the importance of the previous point,
      for the convenience of the reader I report here 
      verbatim the list of misunderstandings appearing in the
      Wikipedia at the {\bf end of 2011}\,\cite{Badmath},\footnote{As
      it is well known, the content of Wikipedia
      is variable with time. The reason I report here the list
      of misunderstandings as it appeared some years ago, and 
      as it has been more ore less until the beginning of 2016 
      -- I have no documented records, but I have been checking it 
      from time to time, in occasion of seminars and courses and I had
      not realized major changes, like the reductions of the items from
      7 to 5 -- is that the present version has been clearly 
      being influenced by the ASA statement of  March 2016.
      (I report here all seven items, although I have to admit that I get 
      lost after the third one -- but you for you seven are still not
      enough see \cite{StevenGoodman})
} 
      highlighting the sentences that mostly concern our 
      discourse.
\begin{quote}
{\small 
\begin{enumerate}
\item
``\,{\bf The p-value is not the probability that the null hypothesis is true.}
      In fact, frequentist statistics does not, and cannot, 
attach probabilities to hypotheses. Comparison of Bayesian 
and classical approaches shows that a p-value can be very close 
to zero while the posterior probability of the null is very close 
to unity (if there is no alternative hypothesis with a large
 enough a priori probability and which would explain the results 
more easily). This is the Jeffreys-Lindley paradox.
\item 
{\bf The p-value is not the probability that a finding is 
     ``merely a fluke.''}
      As the calculation of a p-value is based on the assumption that 
a finding is the product of chance alone, it patently cannot also 
be used to gauge the probability of that assumption being true. 
This is different from the real meaning which is that the p-value 
is the chance of obtaining such results if the null hypothesis is true.
\item The p-value is not the probability of falsely rejecting 
the null hypothesis. This error is a version of the so-called 
prosecutor's fallacy.
\item The p-value is not the probability that a replicating 
experiment would not yield the same conclusion.
\item $(1 - \mbox{p-value})$ is not the probability of the 
alternative hypothesis being true.
\item The significance level of the test is not determined by the p-value.
      The significance level of a test is a value that should 
be decided upon by the agent interpreting the data before 
the data are viewed, and is compared against the p-value 
or any other statistic calculated after the test has been performed. 
(However, reporting a p-value is more useful than simply saying 
that the results were or were not significant at a given level, 
and allows the reader to decide for himself whether to consider 
the results significant.)
\item The p-value does not indicate the size or importance 
of the observed effect (compare with effect size). 
The two do vary together however -- the larger the effect, 
the smaller sample size will be required to get a significant p-value.''
\end{enumerate}
}
\end{quote}
\item If we want to form our minds about which hypothesis is more or
      less probable in the light of all available information, then 
      we need to base our reasoning on {\em probability theory},
      understood as the {\em mathematics of beliefs}, that is essentially 
      going back to the ideas of Laplace. In particular 
      the updating rule, presently known as the {\em Bayes rule} 
      (or Bayes theorem), should be probably  better called
      {\em Laplace rule}, or at least Bayes-Laplace rule. 
\item The `rule', expressed 
      in terms of the alternative {\em causes} 
      ($C_i$) which could possibly produce the {\em effect} ($E$), 
      as originally done by Laplace,\footnote{This is 
      ``Principle VI``, expounded in simple words 
      in  \cite{Laplace2}, in which
      he calls `principles' 
     the principal rules resulting from his theory. 
      Note also 
      that Eq.\,(\ref{eq:BayesRule1}) requires that hypotheses $C_i$
      form a `complete class' (exhaustive and mutually exclusive), 
      while  Eq.\,(\ref{eq:BayesRule2}) is more general,
      although it might require some care in its application, as pointed
      out in \cite{Fenton_neutral_evidence} 
      $[$\,think e.g. at the hypotheses $H_1=C_1\cap C_2$ and $H_2=C_2$,
      implying: i) $P(H_1) \le P(H_2)\ \forall\, E$\,;
     ii) the calculation of $P(C_1\,|\,E)$ and $P(C_2\,|\,E)$ 
      requires extra information$]$. 
      }  is 
      \begin{eqnarray}
       P(C_i\,|\,E,I) &=& \frac{P(E\,|\,C_i,I)\cdot P(C_i\,|\,I)}
                 {\sum_k P(E\,|\,C_k,I)\cdot P(C_k\,|\,I)}\,.
      \label{eq:BayesRule1}
       \end{eqnarray}
       or, considering also $P(C_j\,|\,E,I)$ and taking the ratio of
        the two 
       {\em posterior probabilities},  
       \begin{eqnarray}
        \frac{P(C_i\,|\,E,I)}{P(C_j\,|\,E,I)} &=& 
         \frac{P(E\,|\,C_i,I)}{P(E\,|\,C_j,I)} \times 
         \frac{P(C_i\,|\,I)}{P(C_j\,|\,I)}\,,
        \label{eq:BayesRule2}
       \end{eqnarray}
      where $I$ stands for the {\em background information}, 
      sometimes implicitly assumed.
\item Important consequences  of this rule -- I like to call them 
      Laplace's teachings\,\cite{Badmath}, because they stem 
      from his ``{\em fundamental principle} of that branch of
the analysis of chance that consists of reasoning a
posteriori from events to causes''\,\cite{Laplace2} --
      are:
      \begin{itemize}
      \item It makes no sense to speak about how the probability
            of $C_i$ changes if:
            \begin{enumerate}
            \item there is no alternative cause $C_j$;
            \item the way how  $C_j$ might produce $E$ is not 
                  properly  modelled,
                  i.e. if $P(E\,|\,C_j,I)$ has not been {\em somehow} 
                  assessed.\footnote{It does not matter if the assessment
                  is done 
                  analytically, numerically, by simulation, or just 
                  by pure subjective considerations --  
                 what is important to understand is that 
                 without the slightest guess on what 
                $P(E\,|\,C_j,I)$  could be, and on how much $C_j$ is 
                more or less believable, you cannot modify your 
                `confidence' on $C_i$, as it will be further reminded in 
                 section  \ref{sec:p-values_BF}.}
             \end{enumerate}
      \item The updating of the probability ratio 
            depends only on the so called {\em Bayes factor}\\ 
            \mbox{}\vspace{-0.4cm}
            \begin{eqnarray}
             \frac{P(E\,|\,C_i,I)}{P(E\,|\,C_j,I)}\,,
             \label{eq:BF}
            \end{eqnarray}
            \mbox{} \vspace{-0.4cm}\mbox{} \\ 
            ratio of the probabilities of $E$ given either
            hypotheses,\footnote{
Eq.\,\ref{eq:BF}
            is also known as ``{\em likelihood} ratio'', but I avoid
            and discourage the use of the 
            `{\em l}-word', being a major source of 
            misunderstanding among 
            practitioners\,\cite{BR,GdA_asymmetric}, who
            regularly use the `{\em l}-function' as pdf of the unknown 
            quantity, taking then (also in virtue of an unneeded `principle') 
            its argmax as  {\em most believable value},
            sticking to it in further `propagations'\,\cite{GdA_asymmetric}. 
            (A recent, important example comes from two reports 
             of the same organization, each using the 
            `{\em l}-word' with two
             different 
             meanings\,\cite{ENFSI_Guideline,ENFSI_Manual}.)
            } 
            and {\it not on the probability of other
            events that have not been observed and 
            that are even less probable than $E$} (upon which
            p-values are instead calculated).
      \item One should be careful not to confuse $P(C_i\,|\,E)$
            with $P(E\,|\,C_i)$, and in general  $P(A\,|\,B)$, 
            with $P(B\,|\,A)$. Or,  moving to continuous variables,
            $f(\mu\,|\,x)$ with $f(x\,|\,\mu)$, where: `$f()$' stands here,
            depending on the contest,
            for a {\it probability function} 
            or for a {\it probability density function} (pdf): 
            $x$ and $\mu$ are symbols for observed quantity and 
            `true' value, respectively, the latter being in fact just 
             the {\em parameter of the model we use to describe the physical 
             world}.
      \item Cause $C_i$ is {\em falsified} by the observation 
            of the event $E$ {\em only if} 
            $C_i$ \underline{cannot} produce it, and   not 
             because of the smallness of  
            $P(E\,|\,C_i,I)$. 
      \item Extending the reasoning to continuous observables (generically 
            called $X$)
            characterized by a pdf
            $f(x\,|\,H_i)$, the probability to observe a value in the
            {\em small} interval $\Delta x$ is $f(x\,|\,H_i)\,\Delta x$.
            What matters, for the comparison of two hypotheses in the light
            of the observation $X=x_m$, is
            therefore
            the ratio of pdf's $f(x_m\,|\,H_i)/f(x_m\,|\,H_j)$, and not 
            the smallness of $f(x_m\,|\,H_i)\,\Delta x$, which tends
            to zero as  $\Delta x \rightarrow 0$. Therefore,
            {\em an hypothesis is}, strictly speaking, {\em falsified}, 
            in the light
            of the observed $X=x_m$, {\em only} if $f(x_m\,|\,H_i)=0$.
      \end{itemize}
\item Finally, I would like to stress that {\em falsificability 
      is not a strict 
      requirement for a theory to be accepted as 
      `scientific'}. 
      In fact, in my opinion a weaker condition is sufficient, 
      which I called {\em testability} in \cite{Vulcano}:
      given a theory $T\!h_i$ and possible observational data
     ${\cal D}$, it should be possible to model 
      $P({\cal D}\,|\,T\!h_i)$ in order to compare it 
      with an alternative theory $T\!h_j$ characterized 
      by   $P({\cal D}\,|\,T\!h_j)\ne P({\cal D}\,|\,T\!h_i)$.\footnote{For
      example String Theory ($ST$) supporters should 
      tell us in what $P({\cal D}\,|\,ST)$ differs from 
      $P({\cal D}\,|\,SM)$ from Standard Model, with 
      ${\cal D}$ being past, present 
      or future  {\em observational data}.} 
      This will allow to rank theories in probability in the light 
      of empirical data and of any other criteria, like 
      simplicity or aesthetics\footnote{But we have to be careful 
      with judgments based on aesthetics, which 
      are unavoidably anthropic (and debates on aesthetics 
      will never end, while ancient Romans wisely used to say that
      ``de gustibus non disputandum est'' and, as  
      someone warned, ``if you are out to describe the truth, 
      leave elegance to the tailor.''\cite{Elegance-taylor}. 
      This is more or less what is going on in Particle Physics 
      in the past years, after that nothing new has been found 
      at LHC besides the highly expected observation of the Higgs boson
      in the final state, with many serious theorists humbling 
      admitting that ``Nature does not seem to share our 
      ideas of {\em naturalness}.''} 
      without the requirement of falsification, that cannot be achieved,
      logically speaking,
      in most cases.\footnote{Think for example at all infinite 
      numbers of Gaussian models ${\cal N}(\mu,\sigma)$
      that might have produced the observation $x_m=4$. Since, 
      strictly speaking, any Gaussian might produce any real value, 
      it follows none of the $\infty^2$ models can be falsified. 
      Nevertheless, every one will agree that $x_m=4$ it is {\em more likely} 
      to be attributed to model ${\cal N}(3,1)$ than 
      ${\cal N}(20,1)$. But you cannot say that the observation 
       $x_m=4$ falsifies model ${\cal N}(20,1)$! 
       \label{fn:falsification_gaussians}}
\end{itemize}

\section{ASA statement on statistical significance and p-values}
\subsection{{\em Ante factum}}
The statement of the American Statistical Association on March this
year did not arrive completely unexpected. 
Many scientists were in fact aware and worried of the
``science’s dirtiest secret'', i.e. that
``the ‘scientific method’ of testing hypotheses 
by statistical analysis stands on 
a flimsy foundation''\cite{ScienceNews_2010}. 
Indeed, as Allen Caldwell of MPI Munich  eloquently puts it
(e.g. in \cite{Caldwell_Stellenbosch}) ``The real problem 
is not that people have difficulties in understanding
Bayesian reasoning. The problem is that they do not understand 
the frequentist approach and what can be concluded from 
a frequentist analysis. What is not understood, or forgotten, 
is that the frequentist analysis relates only to possible data 
outcomes within a model context, and not probabilities 
of a model being correct.  
This misunderstanding leads to faulty conclusions.''
 
\begin{figure}[t]
\begin{center}
\epsfig{file=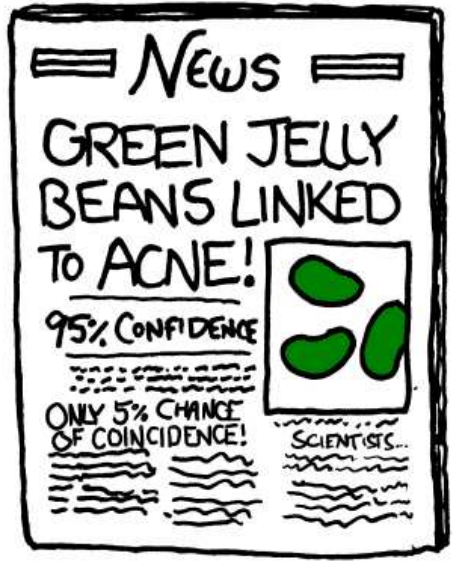,width=0.42\linewidth}
\end{center}
\mbox{}\vspace{-1.0cm}
\caption{\small \sf A `significant' result obtained {\em provando
e riprovando}\,\cite{xkcd-significant}.}
\label{fig:xkcd-significant}
\end{figure}

Faulty conclusions based on p-values are countless
in all fields of research, and frankly I am 
personally much more worried 
when they might affect our 
health\footnote{See e.g.  \cite{Naik,Iorns,Jump_Aug,Jump_Sep}
(for instance Elisabeth Iorns' {\em comment} on 
New Scientist\,\cite{Iorns} 
reports that ``more than half of biomedical findings 
cannot be reproduced'' and ``pharmaceutical company Bayer says it fails to replicate two-thirds of published drug studies'' --\,{ !!!}).} and security, 
 or the future of our planet, rather
then when they spread around unjustified 
claims
of revolutionary discoveries or of possible failures
of the so called 
Standard Model of Particle 
Physics\,\cite{Badmath}.\footnote{Frankly I do not think
that these claims hurt fundamental physics, which I consider
quite healthy and (mostly) done by honest researchers. In fact, false 
alarms might even have positive effects inside the community, 
because they stimulate discussions on completely new
possibilities and encourage new researches to be undertaken, 
as also recognized in the bottom line of de Rujula's cartoon
of Fig. \ref{fig:deRujula_cemetry}. My worries mainly concern
negative reputation the field risks to gain and, perhaps 
even more,  bad education provided to young people,
most of which  will leave pure research and will
try to apply elsewhere the analysis methods they learned
in searching for new particles and new phenomena.} 
For instance,  ``A lot of what is published is incorrect'' reported 
last year {\em The Lancet}'s Editor-in-Chief 
Richard Horton\,\cite{TheLancet}.
This could be because, {\em looking around more or less `at random'},
statistical `significant results' will soon or later  show up  
(as that of the last frame of an {\em xkcd} cartoon 
shown in Fig.\,\ref{fig:xkcd-significant} -- 
see \cite{xkcd-significant} for the full story);
or because dishonest (or driven by wishful thinking, which in Science
is more or less the same) researchers might do some 
{\em p-hacking} (see e.g.  \cite{Nelson} and  \cite{Charpentier})
in order to make 
`significant effects' appear
-- remember that ``if you torture the data long enough, 
it will confess to 
anything''\,\cite{tortured_data}.

A special mention deserves the February 2014 editorial 
of David Trafimow, Director of 
{\em Basic and Applied Social Psychology} (BASP),
in which he takes a strong position
 against 
``null hypothesis significance testing procedure (NHSTP)''
because it ``has been shown to be logically invalid and to provide
little information about the actual likelihood of either the
null or experimental hypothesis''\,\cite{BASP_2014}. 
In fact a large echo (see e.g. 
\cite{Science-Based_Medicine},
\cite{Andrew2015} and 
\cite{ISBA_March2015})
had last year a second
editorial, signed together with his Associate Director
Michael Marks published
on February 15, 2015, in which they announce 
that, after ``a grace period allowed to authors'', 
``from now on, BASP is banning the NHSTP''\,\cite{BASP_2015}.
 
\subsection{{\em Principia}}\label{ss:Principia}
Moving finally to the content of the ASA statement, 
after a short introduction, in which it is recognized 
that ``the p-value $[$\ldots$]$ is commonly misused 
and misinterpreted,'' 
and a reminder of what a p-value 
``informally''' is (``the probability under a specified 
statistical model that a statistical summary of the data 
$[$\ldots$]$would be equal
to or more extreme than its observed value'') a list
of six items, indicated as ``principles'', follows 
(the highlighting is original).
\begin{quote}
{\small
\begin{enumerate}
\item {\bf {\em P}-values can indicate how incompatible the data are
with a specified statistical model.} \\
     \mbox{}\hspace{0.3cm} A $p$-value provides one approach to summarizing
     the incompatibility between a particular set of data and
     a proposed model for the data. The most common
context is a model, constructed under a set of assumptions, 
together with a so-called ``null hypothesis.'' Often
the null hypothesis postulates the absence of an effect,
such as no difference between two groups, or the absence
of a relationship between a factor and an outcome. The
smaller the $p$-value, the greater the statistical 
incompatibility of the data with the null hypothesis, 
if the underlying assumptions used to calculate the $p$-value hold. This
incompatibility can be interpreted as casting doubt on
or providing evidence against the null hypothesis or the
underlying assumptions.
\item {\bf {\em P}-values do not measure the probability 
that the studied hypothesis is true, or the probability that the data
were produced by random chance alone.} 
 \\    \mbox{}\hspace{0.3cm} Researchers often 
wish to turn a $p$-value into a statement about the truth of a null hypothesis, 
or about the probability that random chance produced the observed
data. The $p$-value is neither. It is a statement about data
in relation to a specified hypothetical explanation, and is
not a statement about the explanation itself.
\item {\bf Scientific conclusions and business or policy decisions
should not be based only on whether a {\em p}-value passes
a specific threshold.}
 \\    \mbox{}\hspace{0.3cm} Practices 
that reduce data analysis or scientific inference to mechanical 
``bright-line'' rules 
(such as ``$p < 0.05$'') for justifying scientific 
claims or conclusions can
lead to erroneous beliefs and poor decision making. A
conclusion does not immediately become ``true'' on one
side of the divide and ``false'' on the other. Researchers
should bring many contextual factors into play to derive
scientific inferences, including the design of a study,
the quality of the measurements, the external evidence
for the phenomenon under study, and the validity of
assumptions that underlie the data analysis. Pragmatic
considerations often require binary, ``yes-no'' decisions,
but this does not mean that $p$-values alone can ensure
that a decision is correct or incorrect. The widespread
use of ``statistical significance'' (generally interpreted as
$p \le 0.05$'') as a license for making a claim of a scientific
finding (or implied truth) leads to considerable distortion 
of the scientific process.
\item {\bf Proper inference requires full reporting and transparency}
 \\    \mbox{}\hspace{0.3cm} $P$-values and related analyses 
should not be reported selectively. Conducting multiple analyses of the data
and reporting only those with certain $p$-values 
(typically those passing a significance threshold) renders the
reported $p$-values essentially uninterpretable. Cherry-picking 
promising findings, also known by such terms as
data dredging, significance chasing, significance questing, 
selective inference, and ``$p$-hacking,'' leads to a
spurious excess of statistically significant results in the
published literature and should be vigorously avoided.
One need not formally carry out multiple statistical tests
for this problem to arise: Whenever a researcher chooses
what to present based on statistical results, valid 
interpretation of those results is severely compromised if
the reader is not informed of the choice and its basis.
Researchers should disclose the number of hypotheses
explored during the study, all data collection decisions,
all statistical analyses conducted, and all $p$-values computed. 
Valid scientific conclusions based on $p$-values and
related statistics cannot be drawn without at least knowing 
how many and which analyses were conducted, and
how those analyses (including $p$-values) were selected for
reporting.
\item {\bf A {\em p}-value, or statistical significance, does not measure
the size of an effect or the importance of a result.}
 \\    \mbox{}\hspace{0.3cm} Statistical 
significance is not equivalent to scientific, 
human, or economic significance. Smaller $p$-values
do not necessarily imply the presence of larger or
more important effects, and larger $p$-values do not
imply a lack of importance or even lack of effect. Any
effect, no matter how tiny, can produce a small $p$-value
if the sample size or measurement precision is high
enough, and large effects may produce unimpressive
$p$-values if the sample size is small or measurements
are imprecise. Similarly, identical estimated effects will
have different $p$-values if the precision of the estimates
differs.
\item {\bf By itself, a {\em p}-value does not provide a good measure of
evidence regarding a model or hypothesis.}
 \\    \mbox{}\hspace{0.3cm} Researchers should recognize 
that a $p$-value without
context or other evidence provides limited information.
For example, a $p$-value near 0.05 taken by itself offers only
weak evidence against the null hypothesis. Likewise, a
relatively large $p$-value does not imply evidence in favor
of the null hypothesis; many other hypotheses may be
equally or more consistent with the observed data. For
these reasons, data analysis should not end with the calculation 
of a $p$-value when other approaches are appropriate and feasible.
\end{enumerate}
}
\end{quote}
These words sound as an admission of failure 
of much of the statistics teaching and practice 
in the past {\em many} decades. 
But yet I find their courageous statement still 
somehow unsatisfactory, and, in particular, 
the first principle is in my opinion 
still affected by the kind of `original sin' 
at the basis of p-value misinterpretations and misuse. 
Many practitioners consider in fact a value occurring several 
(but often just a few)
standard deviations from the `expected value' (in the probabilistic
sense) to be a 'deviance' from the model, which is clearly
absurd: no value a model can yield can be considered an 
{\em exception} from the model itself
(see also footnote \ref{fn:falsification_gaussians} -- 
the reason why ``p-values {\em often work}'' 
will be discussed in section \ref{sec:p-values_BF}).
Then, moving to principle 2, 
it is not that ``researchers {\em often wish} to turn a $p$-value 
into a statement about the truth of a null hypothesis''
(italic mine),
as if this would be an extravagant fantasy: reasoning 
in terms of degree of belief of whatever is uncertain
is connatural to the 
`human understanding'\,\cite{Hume}: 
{\em all methods that do not tackle straight the fundamental
issue of the probability of hypotheses, 
in the problems in which this is the crucial question,
are destinated to fail, and to perpetuate misunderstanding
and misuse}.

\section{The `Monster' blessed by the 5 sigmas}
\begin{figure}[t]
\begin{center}
\epsfig{file=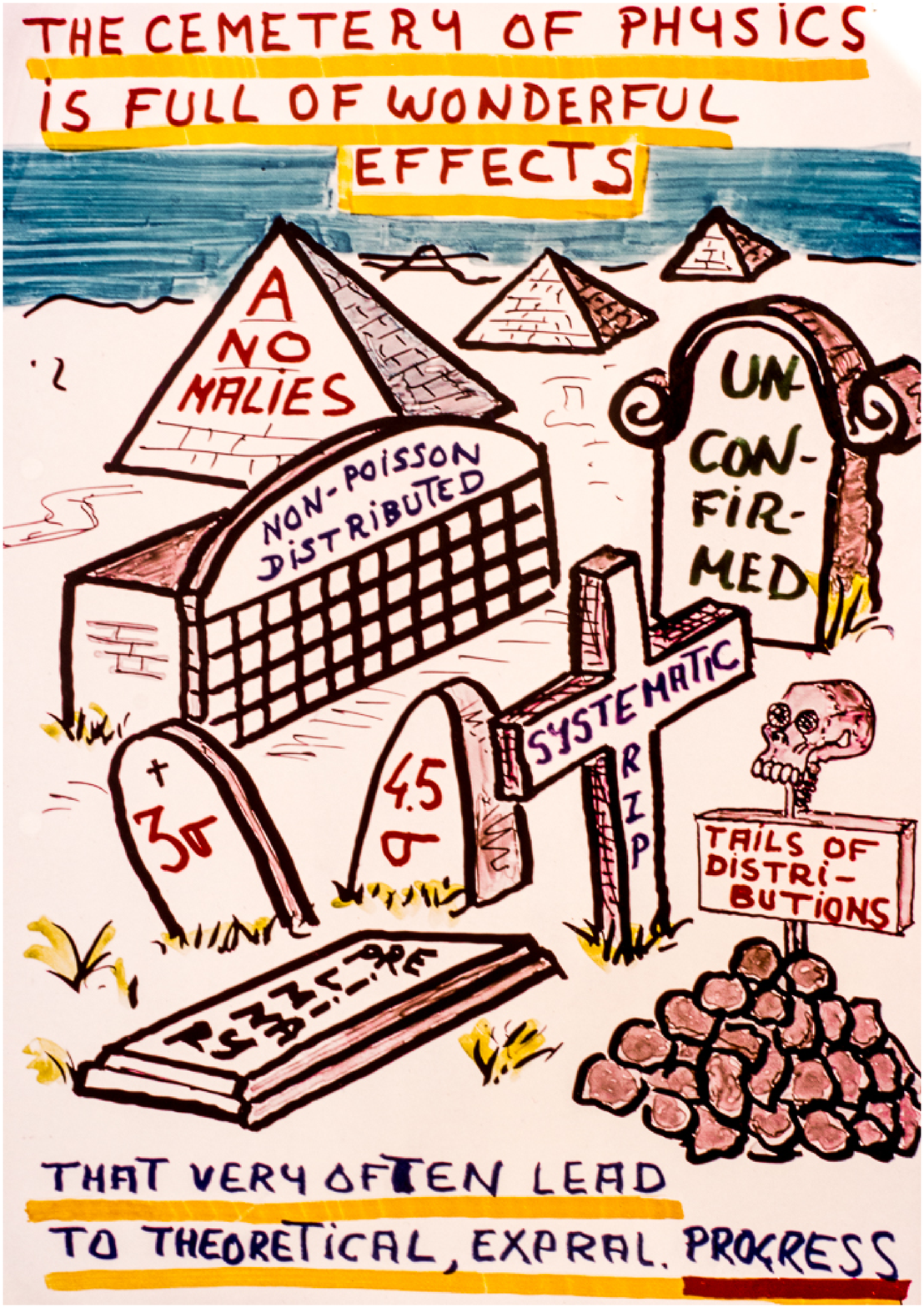,width=0.45\linewidth}
\end{center}
\mbox{}\vspace{-0.8cm}
\caption{\small \sf Alvaro de Rujula's 
Cemetery of Physics\,\cite{deRujula}, with graves indicating 
`false alarms' in frontier physics,
 and not old physics ideas faded out with time, 
like epicycles, phlogiston or aether.}
\label{fig:deRujula_cemetry}
\end{figure}

Rumors that the LIGO interferometers had most likely 
detected a gravitation wave (GW) were circulating
in autumn last year. Personally, the direct information
I got quite late, at the beginning of December, was 
``we have seen a {\em Monster}'', without further detail. 
Therefore, when a few days before February 11 
quantitative rumors talked of 5.1 sigmas, I was disappointed
and highly puzzled.
How could a Monster have {\em only} just a bit more than 
five sigmas? Indeed in the past decades we have seen
in Particle Physics 
several effects of similar statistical significance 
coming and going, 
as Alvaro de Rujula
depicted already in 1985 in his famous
{\em Cemetery of Physics} of 
Fig.\,\ref{fig:deRujula_cemetry}\,\cite{deRujula}.\footnote{Finally
he humorously summarized his very long experience in the 
`{\em de Rujula paradox'}  \cite{deRujula_paradox}:
\begin{quote}
{\sl \bf
If you {\em disbelieve} every result presented as having a 3 sigma, \\ 
or `equivalently' a 99.7\% chance of being correct, \\ 
you will turn out to be {\em right} 99.7\% of the times.
}
\end{quote}
(`Equivalently' within quote marks is 
de Rujula's original, because he knows 
very well that there is no equivalence at all.) 
\label{fn:deRujula_paradox}
} 
Therefore for many of us a five-sigma effect
would have been something worth discussions 
or perhaps further investigations but certainly not a 
Monster.\footnote{``And the July 2012 5-sigma Higgs boson?'', 
you might argue. Come on! That \underline{was} the Higgs boson, 
the highly expected missing tessera to give sense 
to the amazing mosaic of the Standard Model, 
 whose mass had already been somehow inferred 
from other measurements, 
although with quite large uncertainty 
(see e.g. \cite{GdA-GD,GdA-GD_2}).
For this reason the 2011 data were sufficient to many 
who had followed this physics since years (and not sticking
to the 5-sigma dogma) to be highly confident that the Higgs
boson was finally observed in a final state 
diagram\,\cite{Badmath}. Instead, some of those who were 
casting doubt on the possibility of observing the Higgs are the same 
who were giving credit to the December 2015 $\gamma\gamma$
750\,GeV excess at LHC 
(and some even to the Opera's superluminar neutrinos!). 
I hope they will learn from the double/triple 
lesson.\label{fn:5sigmaHiggs}
}
This impression was very evident 
from the reaction many people had {\em after}
seeing the wave form.
``Came on, this is not a five-sigma
effect'', commented several colleagues, more or less using
the same words, ``these are {\em hundreds}
of sigmas!'', a colored expression to say that 
just by eye the hypothesis
Noise was beyond any imagination.\footnote{And indeed we have
also learned that the only serious alternative hypothesis 
taken into account and investigated in detail was 
that of a sabotage!} 

The reason of the `monstrosity' of GW150914 was indeed in Table 1
of the accompanying paper on 
{\em Properties of the binary black hole merger 
GW150914}\,\cite{LIGO-WG150914_properties}: a Bayes factor 
``BBH merger'' Vs ``Noise''\footnote{To be precise,
the competing hypotheses are ``\mbox{BBH-merger\,\&\,Noise}'' Vs 
 ``only Noise''.}
of about $5\times 10^{125}$ (yes, {\em five times 
ten to one-hundred-twenty-five}). This means that, no matter
how small the odds in favor of a BBH merger were  
and even casting doubt on the evaluation 
of the Bayes factor,\footnote{At 
this point a  `technical' remark is in order, 
which is indeed also conceptual and sheds some light on the 
difficulty of the calculation and possible uncertainties on 
the resulting value.  
Given the hypotheses
$H_0$ and $H_1$ and data  ${\cal D}$, the 
Bayes factor $H_1$ Vs $H_0$ is
\begin{eqnarray*}
\frac{P({\cal D}\,|\,H_1,I)}{P({\cal D}\,|\,H_0,I)}\,, 
\end{eqnarray*} 
where for sake of simplicity we identify  $H_1$ with
``BBH merger'' and $H_0$ with ``Noise''. 
Now the question is that 
{\em there is not a single, precisely defined, hypothesis ``BBH merger''}.
And the same is true also for the `null hypothesis' ``Noise''. 
This is because each hypothesis comes with free parameters.
For example, in the case of ``BBH merger'', 
the conditional probability of  ${\cal D}$  
depends on the masses of the two black holes ($m_1$ and $m_2$), 
on their distance
from Earth ($d$) and so on, i.e. 
$P({\cal D}\,|\,H_1,m_1,m_2,d,\ldots,I)$.
The same holds for the Noise, because there is no such a thing 
as ``the Noise'', but rather a noise model with many parameters
obtained monitoring the detectors. So in general, for the 
generic hypothesis $H$ we have 
\begin{eqnarray*}
P({\cal D}\,|\,H,\underline{\theta},I)\,,
\end{eqnarray*}
in which $\underline{\theta}$ stands for the set of parameters 
of the  hypothesis $H$. 
But what matters for the calculation
of the Bayes factor is $P({\cal D}\,|\,H,I)$, and this 
can be evaluated from probability theory taking 
account all possible values of the set of parameters 
$\underline\theta$, weighting them by the pdf 
 $f(\underline\theta\,|\,H,I)$, i.e. `simply' as 
\begin{eqnarray*}
P({\cal D}\,|\,H,I) &=& \int_{\underline{\Theta}}\!
                            P({\cal D}\,|\,H,\underline{\theta},I)\,
                            f(\underline\theta\,|\,H,I)
\,d\underline{\theta}\,.
\hspace{2cm}(F.1)
\end{eqnarray*}
But the game can be not simple at all,
because i) this integral can be very
difficult to calculate; ii) the result, and then the BF, 
depends on the prior $ f(\underline\theta\,|\,H,I)$ about the parameters, 
which have to be properly modeled from the physics case.
A rather simple example, also related to gravitational waves, 
is shown in \cite{BF_con_Pia_e_Sabrina}
and helped dumping down claims of 
GW detection based on p-values, resulting in fact in ineffective
Bayes factors Signal Vs Noise of the order of the unity, 
with values depending on the model considered.
The calculations of the BF's published by 
the LIGO-Virgo Collaboration are {\em much} more complicate than 
those of \cite{BF_con_Pia_e_Sabrina} 
(see \cite{LIGO-WG150914_properties} and \cite{LIGO-June2016}
and references therein, 
in particular \cite{Veitch-Vecchio}), and they have
highly benefitted of Skilling's 
{\em Nested Sampling} algorithm\,\cite{Skilling_NS}. And, for 
the little I can understand 
of BBH mergers, the priors on the parameters appear to have been chosen
safely, so that the resulting BF's seem very reliable. 
\label{fn:nota_BF}
}  
the posterior odds
would be extraordinary large, the probability of 
noise being smaller than Shakespeare's drop of water 
identically recovered from the sea.%
\footnote{
William Shakespeare, {\em The Comedy of Errors}:
\begin{quote}
{\sl 
For know, my love, as easy mayst thou fall \\
A drop of water in the breaking gulf,\\
And take unmingled thence that drop again,\\
Without addition or diminishing,
}
\end{quote}
\mbox{}\vspace{-0.4cm}
}

\section{Cinderella and her sisters}\label{sec:Cinderella}
\begin{figure}[t]
\begin{center}
\epsfig{file=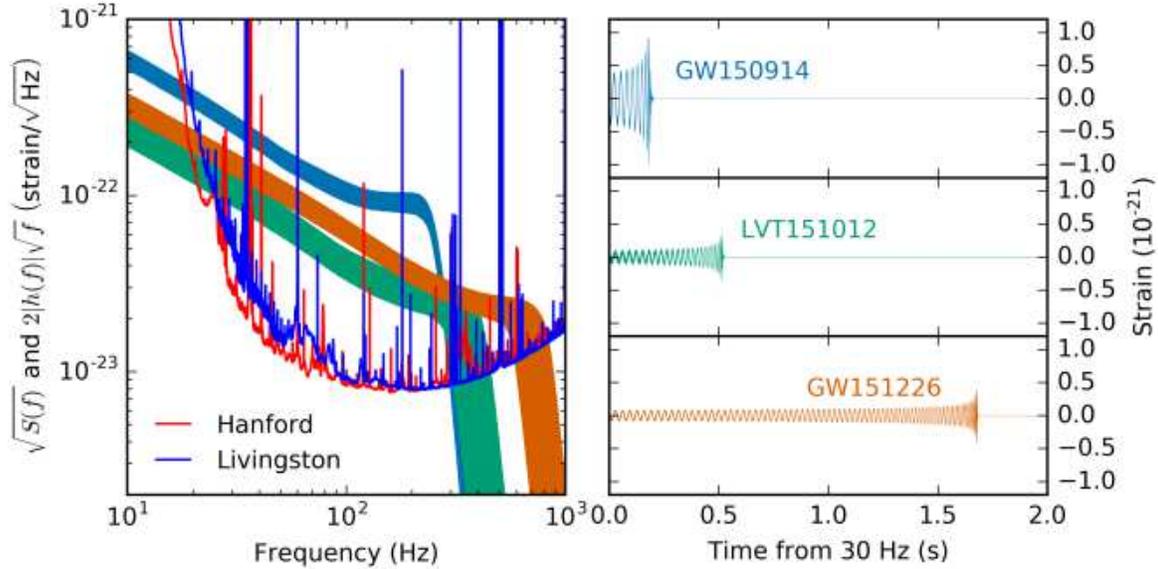,width=\linewidth}
\end{center}
\caption{\small \sf The Monster (GW150914), Cinderella (LVT151012) 
and the third sister (GW151226), visiting us in 2015 
(Fig. 1 of \cite{LIGO-June2016} -- see text for the reason 
of the names). The published `significance' of the three events 
(Table 1 of \cite{LIGO-June2016}) is, 
in the order, ``$> 5.3\,\sigma$'',  ``$1.7\,\sigma$''
and ``$> 5.3\,\sigma$'', corresponding to the 
following p-values: $7.5\times 10^{-8}$, $0.045$,
$7.5\times 10^{-8}$.
The log of the Bayes factors
are instead (Table 4 of \cite{LIGO-June2016}) approximately 
289, 23 and 60, corresponding to Bayes factors 
about $3\times 10^{125}$, $ 10^{10}$ and $ 10^{26}$.} 
\label{fig:Fig1_LIGO-June2016}
\end{figure}

The results of the full observing run 
of the Advanced LIGO detectors
(September 12, 2015, to January 19, 2016)  
have been presented on June 8\,\cite{LIGO-June2016},
slightly updating some of the February's digits.  
Figure  \ref{fig:Fig1_LIGO-June2016} summarizes 
detector performances and results, with some
important numbers (within this context) reminded in the caption. 

The busy plot on the left side shows the sensitivity curves of the 
two interferometers (red and blue curves, with plenty of resonant
peaks) and how the three signals fall inside them (bands with colors
matching the wave forms of the right plot). 
In short, the two curves tell us 
that a signal of a given frequency can 
be distinguished from the noise if its amplitude is above 
them. Therefore all initial parts of the waves, 
when the black holes begin to spiral around each 
other at low frequency, are unobservable, and the bands 
below $\approx 20\,$Hz are extrapolations from the physical models.  
Later, when the frequency increases, the wave enters the 
sensitivity range,\footnote{In analogy, 
imagine someone communicating to us
using an audio signal, whose frequency changes with time,  
from infrasounds to ultrasounds. We can ear the signal only 
when it is in the acoustic region, conventionally in the 
range between 20 and 20,000 Hz, although depending from person 
to person. And, since this sensitivity window is not sharp,
close to its edges loud sounds are better eared
than quiet ones. 
\label{fn:acustic_analogy}} 
which extends up to a given frequency, after which we `loose' it.
The lower and upper boundary frequencies 
depend on the amplitude of the signal, as it also happens
in acoustics. 

The plot on the right shows  finally the `waves'\footnote{To 
be more precise, these {\em are not} data points, but rather the
`adapted filters' that best match them, and therefore they could 
provide a too optimistic impression of what has really being detected. 
Therefore we have to use the Bayes factors provided by the collaboration,
rather than intuitive judgement based on these wave forms.} 
from the instant 
they enter the optimal 30\,Hz sensitivity region 
(the acoustic analogy depicted in 
footnote\,\ref{fn:acustic_analogy} might help):
\begin{itemize}
\item
The wave indicated by GW150914 (the `Monster', with 
GW standing for gravitational wave and 150914 for the
detection date, September 14, 2015) is characterized 
by high amplitude, but short duration in the sensitivity
region, because it fades out at a few hundred hertz. 
\item
GW151226 instead, although of smaller intensity, has a longer `life' 
(about 1.7 seconds)
in the `audible' spectrum, and therefore the signature 
of a BBH merger is also very recognizable.
\item
Then there is the October 12 event, 
LVT151012, which has an amplitude comparable to that of GW151226, but 
smaller duration. 
It has, nevertheless,  about 20 oscillations in 
the sensitivity region, an information that, combined
with the peculiar shape of the signal (remarkably the
crests get closer as time passes, while the amplitude
increases, until something `catastrophic' 
seems to happen) and the fact
that two practically `identical' and `simultaneous' 
signals have been observed by 
the two interferometers 3000 km apart, 
makes the experts highly confident that this is also 
a gravitational wave. 
\end{itemize}
However, even if at a first sight it does not look dissimilar from 
GW151226 (but remember that the waves in 
Fig.\,\ref{fig:Fig1_LIGO-June2016} do not show raw data!),  
the October 12 event,
hereafter referred as 
{\em Cinderella},
is not ranked as GW, but, more modestly, 
as LVT, for LIGO-Virgo Trigger. The reason of the downgrading
is that {\em `she' cannot wear a ``$> 5\sigma$'s dress''
to go together with the `sisters' to the `sumptuous ball 
of the Establishment.'} 
In fact Chance has assigned `her' only a poor, 
unpresentable $1.7\,\sigma$
ranking, usually considered in the Particle Physics community not 
even worth a mention in a parallel session of a minor conference
by an undergraduate student.\footnote{Note how 
the quoted p-value of 0.045  associated to it is just 
below the (in-)famous 0.05 ``significance'' 
threshold reminded in the xkcd cartoon of Fig.\,\ref{fig:xkcd-significant}. 
I hope it is so just by chance and that no 
``$\mbox{p-value}\le 0.05$'' requirement was applied to the data, 
then filtering out other possible good signals.} 
But, despite the modest `statistical significance', 
experts are highly confident, 
because of physics reasons\footnote{Detecting 
something that has good reason to exist , 
      because of our understanding of the Physical World 
      (related to a network of other experimental facts 
      and theories connecting them!), 
      is quite different from just observing
      an unexpected bump, 
      possibly due to background, even if with small probability,
     as already commented in footnote \ref{fn:5sigmaHiggs}. 
     $[$And remember that whatever we observe in real life, 
      if seen with high enough resolution
      in the $N$-dimensional phase space, \underline{had} {\em very
      small} probability to occur! 
      (imagine, as a {\em simplified} example,  
      the pixel content of any picture you take walking on the road, 
      in which $N$ is equal to five, i.e two plus the RGB code
      of each pixel).$]$ 
 }
(and of their understanding of background), that this 
is also a gravitational wave radiated by a BBH merger, 
much more than the 87\% quoted in 
\cite{LIGO-June2016}.\footnote{To understand how much people
believe on a scientific statement 
it is often useful, besides proposing bets \cite{Badmath},  
to ask about the  complementary hypothesis. 
For example when I see a 90\% C.L.
upper limit on a quantity, I ask ``do you 
really believe 10\% that the value is 
\underline{above} that limit'',
or, even more embarrassing, ``please use your method to evaluate 
the 50\% C.L. upper limit, then, whatever number comes out,
tell me if you really believe 50-50 that the value could be in 
either side of the limit, and be ready to accept a bet with 1 to 1 odds 
in the direction \underline{I} will choose.'' 
(To learn more about the absurdities of `frequentistic coverage' 
and also about limits derived from `objective Bayesian methods,'
see section 10.7 and chapter 13 of \cite{BR}.) 
In the case of this 87\% probability that LVT151012 
is a GW from  BBH merger the question 
to ask is ``do you really believe 13\%, i.e. about 1 to 7, that 
this event \underline{is not} a gravitational wave due to a BBH merger?''  
(and we should not accept any answer which is, even partially, based
to the smallness of the sigmas.) As a matter of fact 
I find this 87\% beyond my understanding, because such a probability 
has to depend on the prior probability of BBH mergers. For this 
reason I will focus in the sequel only on Bayes factors and 
how they ({\em do not simply}) relate to p-values.}
Indeed the most useful number experimentalists
can provide to the scientific community to quantify 
how the experimental data alone favor the 'Signal' hypothesis
is the Bayes factor, as expounded in the preamble. 
And this factor 
is very large also for Cinderella: $\approx 10^{10}$. 
This means that,
even if your initial odds Signal Vs Noise were one 
to one million, the 
observation of the LIGO interferometers turns them into
10,000 to 1, i.e. a probability of BBH merger of 
99.99\%.\footnote{Note that this probability depends on 
set of hypotheses taken in account. If another, alternative 
physical hypothesis $H^*$ to explain the LIGO signals is 
considered, 
than the Bayes factor of  $H^*$ Vs ``BBH merger'' has to be evaluated,
and the absolute probabilities re-calculated accordingly.} 

Now the question is, how can a modest  $1.7\sigma$ effect
be compatible with a Bayes factor as large as 
 $10^{10}$? The solution to this apparent 
paradox will be given in the next section, but 
I anticipate the answer: {\em p-values and BF's are two different
things, and there is no simple, general rule, inside probability theory,
that relates them.} 

\section{P-values Vs Bayes factors}\label{sec:p-values_BF}
Having discussed at length this topic elsewhere 
(see in particular sections 1.8, and 10.8 of \cite{BR}),
I sketch here the main points, 
with the help of some plots. This is obviously 
a didactic example and does not enter at all into the
(very complicate and CPU time consuming) details
of the analysis of the interferometer data
(see footnote \ref{fn:nota_BF}).
In particular a direct observation will be considered, while
in general hypothesis tests are performed on a {\em statistic} 
chosen with {\em large freedom}.\footnote{It is perhaps
important to remind that, among other problems, 
p-values are affected by arbitrarity of the test variable
used (see e.g. \cite{100tests}), as well by the chosen
subset of data. With some 
experience I have developed my {\em golden rule}: 
\begin{quote}
{\sl \bf The more exotic is the name of
the test, the less believe the result.
}
\end{quote}
The rationale is that I’m pretty sure
that several more common tests have
been discarded before arriving to that which 
provided the desired significance.} 
So  we just consider
here simple models $H_i$ that could produce the 
quantity $x$ according to pdf's $f(x\,|\,H_i,I)$.
\begin{itemize}
\item As reminded above, according to probability theory 
      what matters for the update of relative beliefs
      is the ratio of the pdf's. For example the observation $x_m=5$ 
      shown in the upper plot of Fig.\,\ref{fig:xm_H1-6}
      modifies our beliefs
      in favor of $H_3$, with respect to $H_1$ and $H_2$, 
      {\em no matter the size of the area 
       under the pdf's right of $x_m$}. 
\begin{figure}[!t]
      \begin{center}
      \begin{tabular}{c}
      \epsfig{file=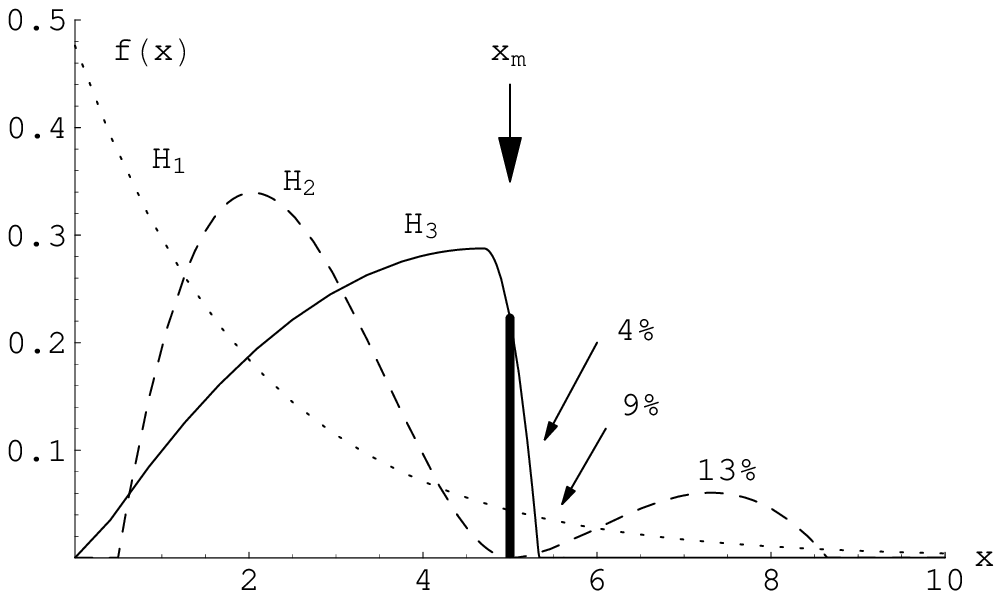,width=0.7\linewidth} \\
       \epsfig{file=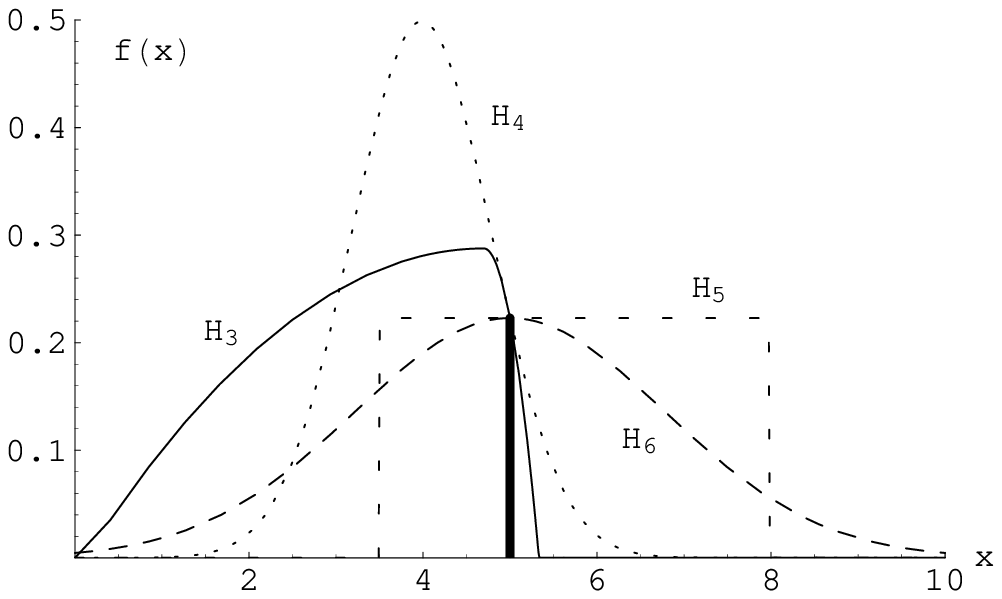,width=0.7\linewidth}
       \end{tabular}
      \end{center}
\caption{\small \sf Several models that could have produced
         the observed value of $x_m$\,\cite{BR}.} 
\label{fig:xm_H1-6}
\end{figure}
\item In particular $H_2$ is ruled out (`falsified') because, 
      being $f(x_m\,|\,H_2)=0$, it cannot produce the observation, 
      despite it provides the highest 
      probability of $X > x_m$.\footnote{Note that, contrary
      to the similar probabilities for the models $H_1$ and $H_3$,
      this 13\% is not a p-value, because 
      $f(x\,|\,H_2) \ge f(x_m\,|\,H_2)\ \forall x > x_m$, while a p-value
      implies an integral on `less probable' values.}
\item It follows that, if the values of pdf's $f(x_m\,|\,H_i)$  
      are equal for all $H_i$, as in the lower plot of Fig.\,\ref{fig:xm_H1-6},
      then the experiment is irrelevant and we 
      hold our beliefs, 
      {\em independently of} how far $x_m$ occurs from the expected 
      values $\mbox{E}[X\,|\,H_i]$, or of the size of the area left or right 
      $x_m$. 
\item The reason why p-values `often work' 
      (and can then be useful {\em alarm bells} when
      getting experiments running, or validating 
      freshly collected data),  
      is quite simple.
      \begin{itemize}
      \item Small p-values are normally associated  
            to small values of the pdf, as shown in the
            upper plot of Fig.\,\ref{fig:xm_p-value}.
\begin{figure}[!t]
      \begin{center}
      \begin{tabular}{c}
      \epsfig{file=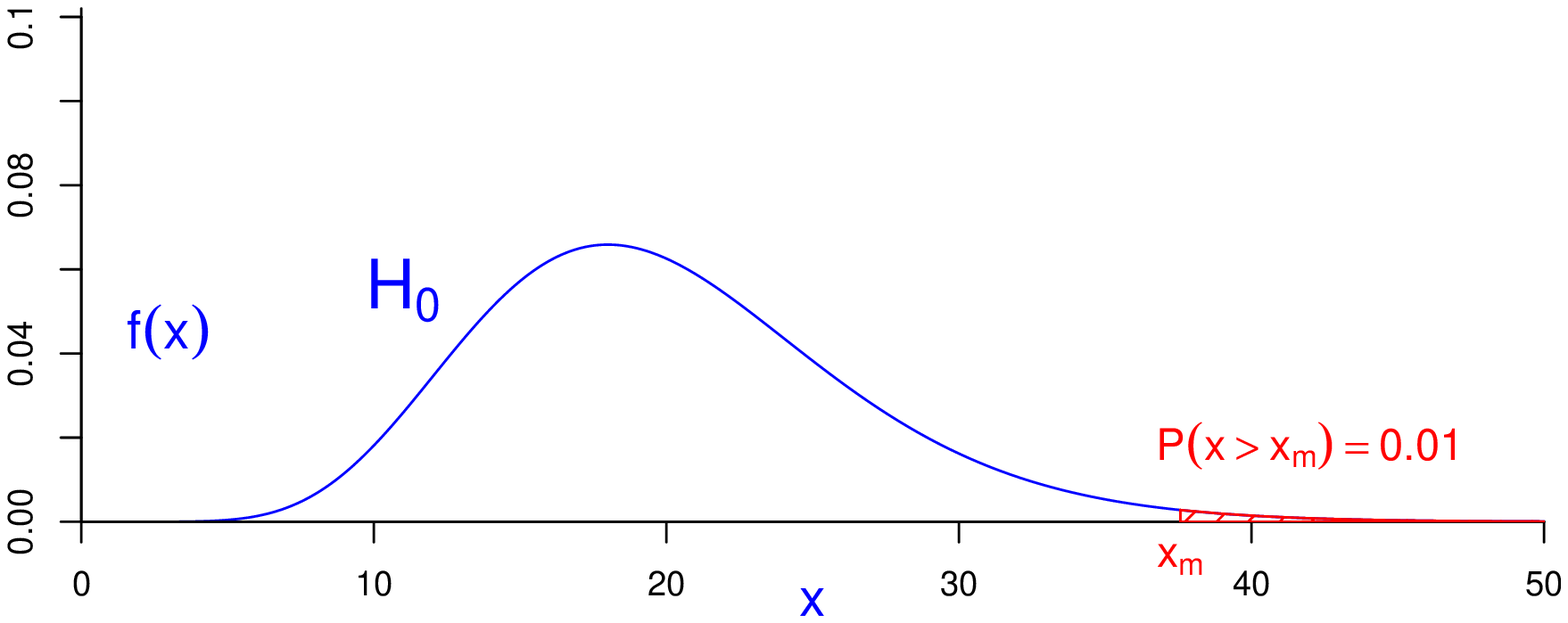,width=0.87\linewidth} \\
       \epsfig{file=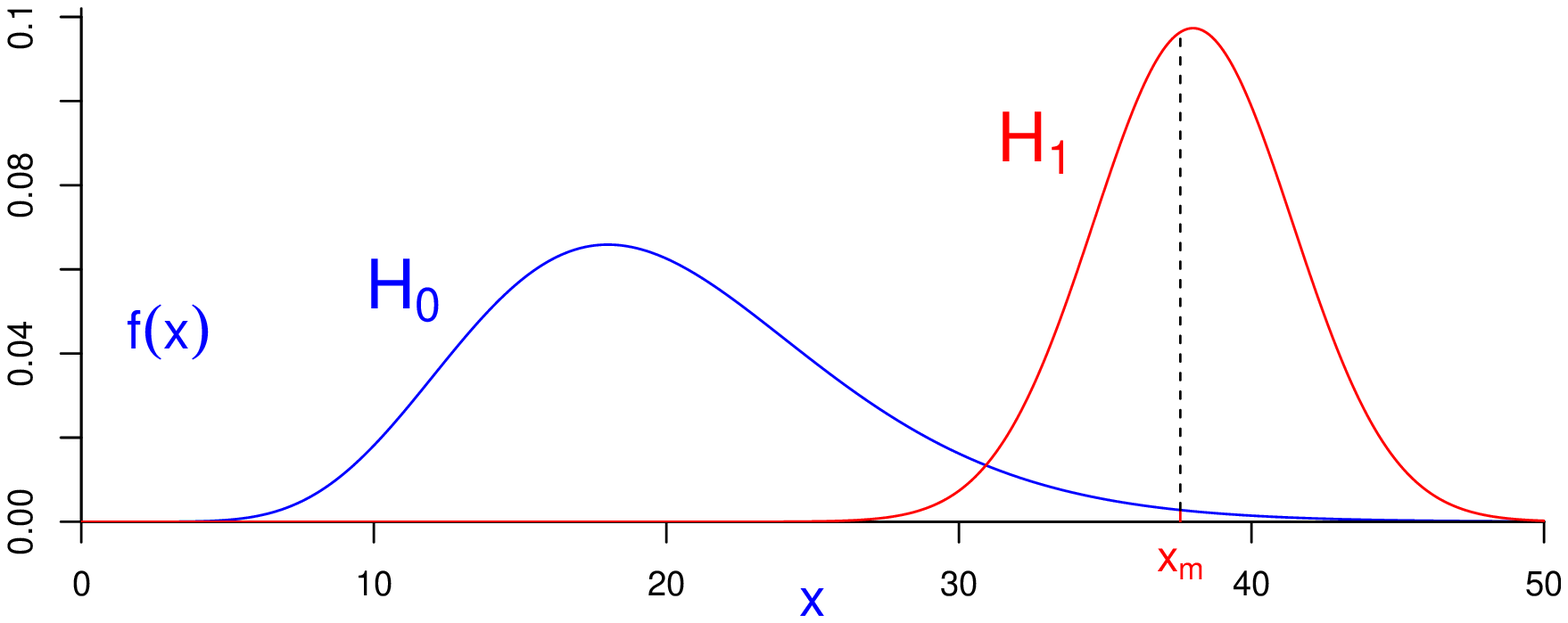,width=0.87\linewidth}
       \end{tabular}
      \end{center}
\caption{\small \sf Pdf's of $X$ given the null hypothesis 
         $H_0$ and the alternative hypothesis $H_1$.} 
\label{fig:xm_p-value}
\end{figure}
      \item It is then {\em conceivable} 
            an alternative hypothesis $H_1$  
            such that $f(x_m\,|\,H_1) \gg f(x_m\,|\,H_0)$,
            as shown in the bottom plot of Fig.\,\ref{fig:xm_p-value}.
      \item Then, {\bf if} this is the case, the observed $x_m$
            {\em would push our beliefs towards} $H_1$, in the sense
            $\mbox{BF}(H_1:H_0) = \frac{f(x_m\,|\,H_1)}
                                        {f(x_m\,|\,H_0)}
              \gg 1      
            $\,.  
      \item  {\bf BUT} we need to take into account also 
             the priors odds $P(H_1\,|\,I)/P(H_0\,|\,I)$.
      \item  In the extreme case such a conceivable $H_1$ 
             \underline{could not exist}, 
             or it could be  
             \underline{not} 
\underline{believable},\footnote{For the distinction
             between what is {\em conceivable} (``Nothing is more free 
             than the imagination of man'') and what is {\em believable} 
             a reference to David Hume\,\cite{Hume} is a must.}
             or it could be just \underline{ad hoc}, as it happens in recent 
             years, with a plethora 
             of `theorists' who give credit to any 
             fluctuation.
             If this is the case, as it is often the case in frontier physics,
             then
            \begin{itemize}
            \item[$\Rightarrow$] $P(H_1\,|\,I)/P(H_0\,|\,I) \rightarrow 0$
            \item[$\Rightarrow$] {\em the smallness of the 
                p-value is irrelevant!}
            \end{itemize} 
      \end{itemize}
      (Note that if, instead of the smallness of the value of the pdf,
       the rational were really the smallness of the area below the pdf, 
       than the absurd situation might arise in which one could choose 
       a ``rejection area'' anywhere, as shown in chapter 1 of \cite{BR}.) 
\item Finally, in order 
      to understand the apparent paradox of 
      large p-value and indeed very large BF, think at a very predictive     
      model $H_1$, whose pdf of the observable $x$ 
      overlaps with that of $H_0$, like in the upper plot of 
      Fig.\,\ref{fig:high_p-value_high_BF}. 
\begin{figure}
      \begin{center}
      \begin{tabular}{c}
       \epsfig{file=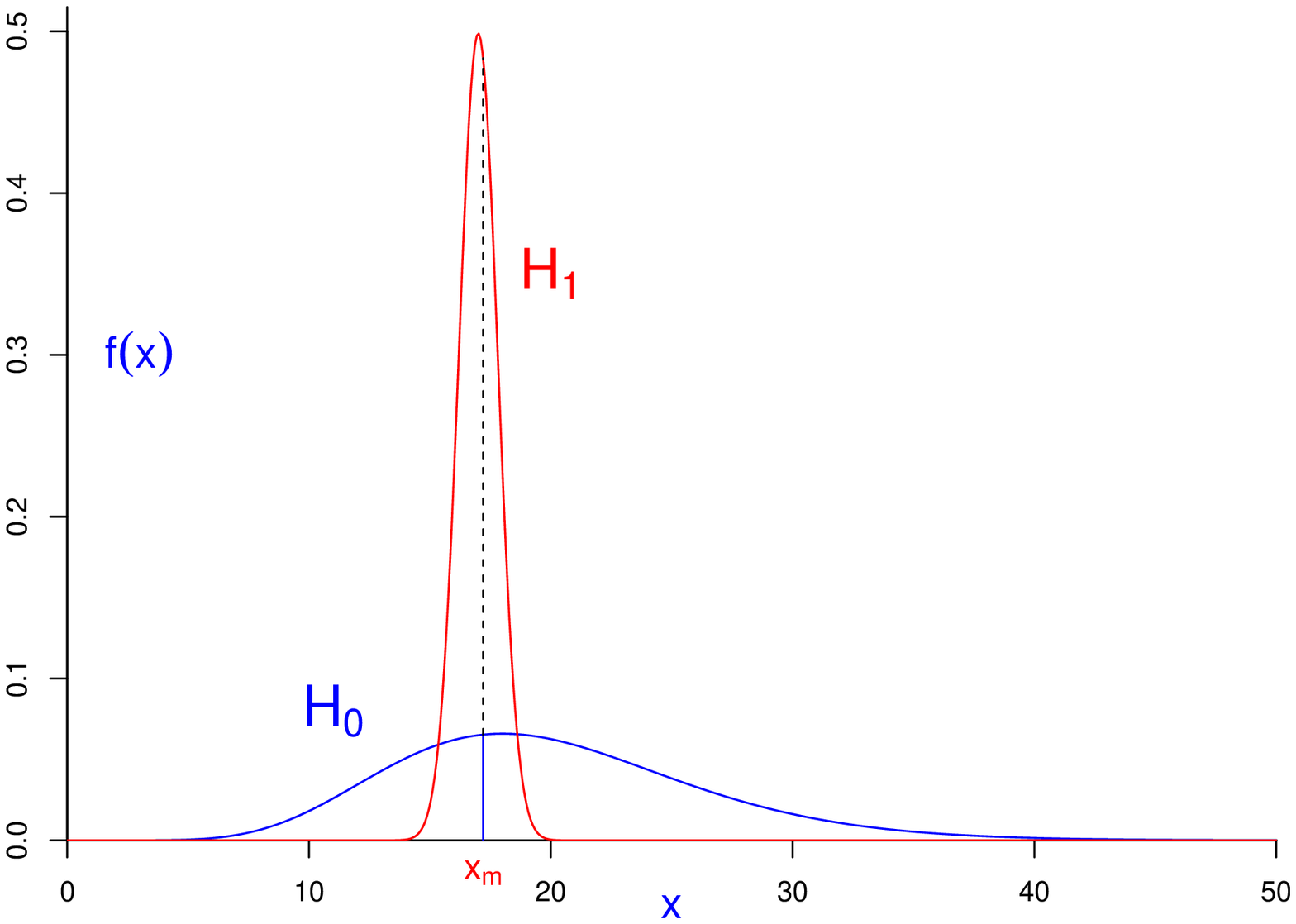,width=0.87\linewidth} \\
        \epsfig{file=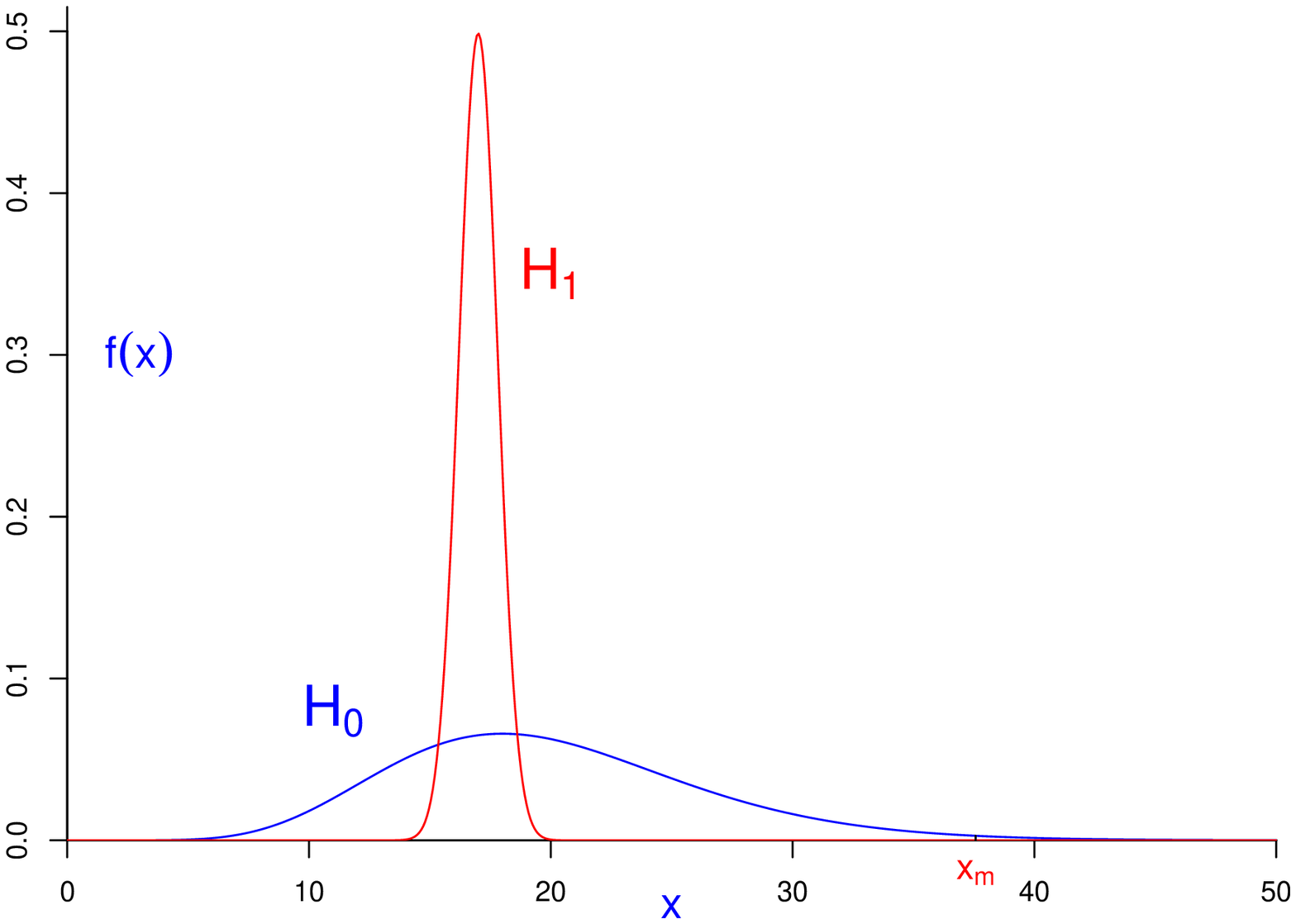,width=0.87\linewidth}
       \end{tabular}
      \end{center}
\caption{\small Pdf's of $X$ given the null hypothesis 
         $H_0$ and the alternative hypothesis $H_1$
         (case of overlapping pdf's).} 
\label{fig:high_p-value_high_BF}
\end{figure}
      We clearly see that  $f(x_m\,|\,H_1) \gg f(x_m\,|\,H_0)$, 
            thus resulting in a Bayes factor highly in favor of $H_1$,
            although the p-value calculated from the null hypothesis
            $H_0$ would be absolutely 
            {\em insignificant}. Something {\em like that}
            occurs in the analysis of the gravitational wave 
            analysis, the case of Cinderella being the most 
            striking one.\footnote{I would like to remind that 
            this is just an academic example to show that 
            effects of this kind are possible and, as far as
            the GW analysis, I rely on the LIGO-Virgo collaboration 
            for the evaluation of p-values and Bayes factors. 
            I am not arguing at all that there could be 
            mistakes in the calculation of the p-values, 
            but rather that it is the interpretation of the latter 
            to be troublesome. Finally, people 
            mostly used to perform $\chi^2$ tests must have
            already realized that the example does not
            apply {\em tout court} to what they do,
            because in that case $H_1$ is usually `richer'
            than $H_0$ and it has then a higher level of adaptability. 
            Therefore the observed value of $\chi^2$ decreases
            (with a `penalty' that frequentists quantify with 
            a reduced number of degree of freedom). As a consequence, the 
            measured value of the test variable is different 
            under the two hypothesis, and, 
            in order to distinguish them, let us indicate the first
            by $\chi^2_0$ and the second by $\chi^2_1$.
             What instead still holds, of the example sketched 
            in the text,
            is that the adaptability of $H_1$ makes 
            the p-value calculated from $f(\chi^2_1\,|\,H_1)$   
            larger that that calculated from 
            $f(\chi^2_0\,|\,H_0)$,
            $$
            \int_{\chi^2_{1_m}}^\infty\!f(\chi^2_1\,|\,H_1)\,d\chi^2_1 \  > \  
            \int_{\chi^2_{0_m}}^\infty\!f(\chi^2_0\,|\,H_0)\,d\chi^2_0\,,  
            $$
             and therefore 
            $H_1$ `gets preferred' to $H_0$. But, as stated in the text,
            the alternative hypothesis $H_1$ could be hardly 
            believable, and therefore its `nice' p-value will not 
            affect the credibility of $H_0$. This almost regularly happens 
            when suspicions against $H_0$ only arise from {\em event counting}
            in a particular variable,
            {\em without any specific physical signature}. 
            $[$As a side remark, 
            I would like to point out, or to remind, that 
            one of the nice features of the Bayes factor calculated
            integrating over the prior parameters of the model, 
            as sketched in footnote \ref{fn:nota_BF}, is that
            models which have a large numbers of parameters,
            whose possible values {\em a priori} extend over
            a large (hyper-)volume, are 
            suppressed by the integral $(F.1)$ with respect  
            to `simpler' models. This effect is known as 
            {\em Bayesian Occam's razor} and is independent from 
            other considerations which might enter in the choice 
            of the priors.
            Those interested to the subject
            are invited to read chapter 28 of David MacKay's great
            book\,\cite{MacKay_book}.$]$} 
\item And `paradoxically' -- this is just a colloquial term,
      since there is no paradox at all --  large 
      deviations from the expected value of $x$ given $H_0$,
      corresponding to  small p-values,  
      are those which favor $H_0$, if $H_1$ and $H_0$ 
      are the only hypotheses in hand, as shown in the
      bottom plot of the same figure. 
      Now, in the light of these examples, 
      I simply re-propose you the following sentence from the first principle
      of the ASA's statement
      {``The smaller the $p$-value, the greater the statistical 
      incompatibility of the data with the null hypothesis, 
      if the underlying assumptions used to calculate the 
      $p$-value hold.''}\,\cite{Asa_statement} 
      As you can now understand, 
      it is not a matter of assumptions concerning $H_0$, 
      but rather on whether alternative hypotheses to $H_0$ are 
      conceivable and, more important, believable!
      \end{itemize}

\subsection{Playing with simulations}
I hope it is now clear the reason why p-values and Bayes factors
have in principle nothing to do with each other, and why 
p-values are not only responsible of unjustified claims 
of discoveries, but might also relegate genuine signals
to the level of fluke, or reduce their `significance',
the word now used as normally understood
and not with the `technical meaning' of statisticians. 
But since I know that many might not be used 
with the reasoning just shown, 
I made a little R script\,\cite{R}, 
so that those who are still sceptical can run it 
and get a feeling of what is going on.
{\small 
\begin{verbatim}
# initialization
mu.H0 <- 0; sigma.H0 <- 1
mu.H1 <- 0; sigma.H1 <- 1e-3
p.H1  <- 1/2
mu    <- c(mu.H0, mu.H1)
sigma <- c(sigma.H0, sigma.H1)

# simulation function
simulate <- function() {
  M <- rbinom(1, 1, p.H1); x <- rnorm(1, mu[M+1], sigma[M+1])
  x <- rnorm(1, mu[M+1], sigma[M+1])
  p.val <- 2 * pnorm(mu[1] - abs(x-mu[1]), mu[1], sigma[1]) 
  BF <- dnorm(x, mu[2], sigma[2]) / dnorm(x, mu[1], sigma[1])
  lBF <- dnorm(x, mu[2], sigma[2], log=TRUE) - dnorm(x, mu[1], sigma[1], log=TRUE)
  cat(sprintf("x = %.5f  =>  p.val = %.2e,  BF = %.2e  [ log(BF) = %.2e ]\n",
              x, p.val, BF, lBF))
  return(M)
}
\end{verbatim}
}
\noindent
By default $H_0$ is simply a standard Gaussian distribution 
($\mu=0$ and $\sigma=1$), while $H_1$ is still a Gaussian
centered in 0, with a very narrow width ($\sigma=1/1000$). 
The prior odds are set at 1 to 1, i.e. $P(H_1)=P(H_0)=1/2$.
Each call to the function {\tt simulate()} prints the values
that we would get in a real experiment 
({\tt x}, p-value, Bayes factor and its log) and returns 
the true model (0 or 1), stored in a vector variable for later check.
In this way you can try to infer what was the real cause
of {\tt x} before knowing  the `truth' 
(in simulations we can, in physics we cannot!). 
Here are the results of a small run, 
with   {\tt x = 12} chosen in order to fill the page,
thus postponing the solution to the next one. \\
\vspace{-0.6cm}
{\footnotesize
\begin{verbatim}
> set.seed(150914); n=12; M <- rep(NA, n); for(i in 1:n) M[i] <- simulate()
x = -0.00079  =>  p.val = 9.99e-01,  BF = 7.29e+02  [ log(BF) = 6.59e+00 ]
x = -0.62293  =>  p.val = 5.33e-01,  BF = 0.00e+00  [ log(BF) = -1.94e+05 ]
x = -0.00029  =>  p.val = 1.00e+00,  BF = 9.57e+02  [ log(BF) = 6.86e+00 ]
x = -0.00162  =>  p.val = 9.99e-01,  BF = 2.68e+02  [ log(BF) = 5.59e+00 ]
x = -0.39258  =>  p.val = 6.95e-01,  BF = 0.00e+00  [ log(BF) = -7.71e+04 ]
x = -0.82578  =>  p.val = 4.09e-01,  BF = 0.00e+00  [ log(BF) = -3.41e+05 ]
x = 0.00073  =>  p.val = 9.99e-01,  BF = 7.69e+02  [ log(BF) = 6.64e+00 ]
x = -0.00012  =>  p.val = 1.00e+00,  BF = 9.93e+02  [ log(BF) = 6.90e+00 ]
x = 0.22295  =>  p.val = 8.24e-01,  BF = 0.00e+00  [ log(BF) = -2.48e+04 ]
x = -0.00022  =>  p.val = 1.00e+00,  BF = 9.76e+02  [ log(BF) = 6.88e+00 ]
x = 0.00117  =>  p.val = 9.99e-01,  BF = 5.07e+02  [ log(BF) = 6.23e+00 ]
x = -1.03815  =>  p.val = 2.99e-01,  BF = 0.00e+00  [ log(BF) = -5.39e+05 ]
\end{verbatim}
}
\noindent
And {\em the winners are}:\\
\vspace{-0.6cm}
{\small 
\begin{verbatim}
> M
 [1] 1 0 1 1 0 0 1 1 0 1 1 0 0 1 1 0 0 0 0 1 0 1 1
\end{verbatim}
}
\noindent
It should not be any longer a surprise that the best figure
to discriminate between the two models is the Bayes factor
and not the p-value.\footnote{If you don't like how the p-value is
calculated in the script, because you might argue about 
one-side or two-sides tail(s), you are welcome to recalculate it, 
but the substance of the conclusions will not change.} 
You can now play with the simulations, varying the parameters. 
If you want to get a situation yielding Bayes factors 
of ${\cal O}(10^{10})$ 
you can keep the standard parameters of $H_0$, 
fixing instead $\mbox{{\tt mu.H1}}$ at $1.7$
and $\mbox{{\tt sigma.H1}}$ at $\approx 4\times 10^{-10}$. 
Then you can choose {\tt p.H1} at wish and run the simulation.
(You also need to change the numbers of digits of {\tt x}, 
replacing ``{\tt \%.5f}'' by ``{\tt \%.11f}'' 
inside {\tt sprintf()}.)

\section{Conclusions}
Uncritical or wishful use of p-values can be dangerous, 
not to speak of unscrupulous p-hacking. While years ago
these criticisms were raised by a minority of thorny Bayesians,
now the effect on the results in 
several fields of science and technology
is felt as a primary issue.\footnote{In the meanwhile
it seems that particle physicists are hard 
in learning the lesson and the number of graves in the
{\em Cemetery of physics} (Fig. \ref{fig:deRujula_cemetry})
has increased since 1985, the last 
{\em funeral} being recently celebrated in Chicago on August 5,
with the following obituary for the {\em dear departed}:
``The intriguing hint of a possible resonance at 750 GeV decaying
 into photon pairs, which caused considerable interest from
 the 2015 data, has not reappeared in the much larger 
2016 data set and thus appears to be a statistical 
fluctuation''\,\cite{Annuncio_morte_750GeV}.
And de Rujula's {\em dictum} (footnote \ref{fn:deRujula_paradox}) 
gets corroborated.
 Someone would argue
that this incident has happened 
because the sigmas were only about three and not 
five. But it is not a question of sigmas, but of Physics, 
as it can be understood by those who in 2012 incorrectly turned the $5\sigma$ 
into 99,99994\% ``discovery probability'' 
for the Higgs\,\cite{GG_CdS_Higgs}, while in 2016 are sceptical  
in front of a  $6\sigma$ claim
(``if I have to bet, my money is on the fact that the result will not
survive the verifications'' \,\cite{GG_CdS_ungheresi}): 
 the famous  
``du sublime au ridicule, il n'y a qu'un pas'' seems really appropriate!
(Or the less famous, outside Italy, ``siamo uomini o caporali!?'')
Seriously, the question is indeed that, now that predictions 
of New Physics
around what should have been
a {\em natural} scale substantially all failed, the only
`sure' scale I can see seems Planck's scale. 
I really hope that LHC will surprise us, 
but hoping and believing are different things.  
And, since I have the impression that are too many nervous people
around, both among experimentalists and theorists, and because 
the number of possible histograms to look at is quite large, 
after the {\em easy bets} of the past years (against CDF peak
and against superluminar neutrinos in 2011; in favor 
of the Higgs boson in 2011; against the 750\,GeV di-photon 
in 2015, not to mention that against Supersymmetry going 
on since it failed to predict new phenomenology
{\em below} the $Z_0$ -- or the $W$? --  mass at LEP, thus inducing me
more than twenty years ago to gave away all SUSY Monte Carlo 
generators I had developed
in order to optimize the performances of the HERA detectors.)$^*$  
I can serenely bet, as I keep saying since July 2012, 
that {\bf the first 5-sigma claim from LHC will be a fluke}.  
(I have instead little to comment on the sociology 
of the Particle Physics 
theory community and on the validity of `objective'
criteria to rank scientific value and productivity, 
being the situation self evident from 
the hundreds of references in a  
review paper which even had in the front page a fake 
PDG entry for the particle\,\cite{Strumia}
and other amenities you can find on the web, 
like \cite{Game of Thrones}.) \label{fn:LHC_750GeV}

$^*$\,{\bf Note added}: on August 22, 2016 a supersymmetry bet among 
theorists has been settled in Copenhagen, {\em declaring winners 
those who betted against supersymmetry}\,\cite{ScommessaHiggsCopenhagen}. 
But I do not think all 
SUSY supporters will agree, because some of them seem to behave like
the guy who said (reference missing) ``I will not die, and nobody
will be able to convince me of the opposite'' -- 
try to convince a dead man he died!
} 
The statement of the American 
Statistical Association is certainly commendable in 
addressing the issue, but it is 
in my opinion unsatisfactory not admitting 
that the question is inherent to all statistical methods
that refuse the very idea of probability of hypotheses, or
of ``probability of causes'', i.e. 
what Poincar\'e used to call
``the essential problem of the experimental method.''

While I had experienced several times in the past, 
including this winter, 
claims of possible breaking discoveries in Particle Physics simply 
due to misinterpretations of p-values, for the first time
I have realized of a case in which judgements based on p-values
strongly reduce the `significance' of important results. 
This happens with the gravitational wave events reported 
this year by the LIGO-Virgo collaboration, and in particular 
with the October 12 events timidly reported as 
a LIGO-Virgo Trigger (`Cinderella'), because of its 1.7 sigmas, in spite
of the huge Bayes factor of about $10^{10}$, that should instead
convince any hesitating physicist about its nature
of a gravitational wave radiated by a Binary Black Hole merger,
especially in the light of the other, more solid two events
(`the two sisters').\footnote{The 
last point deserves a comment, because someone would
object that the three events are ``independent'' and, ``having 
nothing to do with each other, we have to prove
one by one  1) first, that it is a gravitational wave, 
and {\em then } 2) that it comes from BBH merger.'' 
In reality it is consistency of many things, including 
the fact that the values of the inferred parameters 
fall in the expected region, that makes us to 
believe that they are gravitational waves
\underline{and} come from a BBH merger. 
This is because Physics, meant as a Science, 
i.e. an activity of our minds to understand the Physical World,
can be viewed as a large {\em network} of experimental facts
and models, connecting each other 
(``a matrix of beliefs'', as historian Galison
puts it \cite{Galison}). For this reason it is very hard, 
or even impossible, to accommodate in the overall picture 
a new observation that breaks dramatically the net, 
like the 2011 `superluminar neutrinos.' Not by chance 
the title of the February 11 paper was
{\em Observation of Gravitational Waves from a Binary Black Hole Merger}
stressing {\em both observations at once} 
(or if you like `discoveries' -- 
but I don't want to enter into the question of what is `discovery' 
and what is `observation', and I find it commendable
that the collaboration used low profile terminology). 
Therefore, after the first event we 
feel highly confident that events of that kind, with masses 
of that order of magnitude do exist, and with this respect
the three events are not independent, if we refer to probabilistic
independence. More precisely they are 
{\em positively correlated}, i.e. 
\begin{eqnarray*}
P(E_2=\mbox{``BBHm's GW''}\,|\,E_1=\mbox{``BBHm's GW''}, I) &>& 
P(E_2=\mbox{``BBHm's GW''}\,|\,I) \\
P(E_1=\mbox{``BBHm's GW''}\,|\,E_2=\mbox{``BBHm's GW''}, I) &>& 
P(E_1=\mbox{``BBHm's GW''}\,|\,I)\,,
\end{eqnarray*}
and so on. This effect, indeed rather intuitive, 
can been shown to occur in a quantitative way, modelling 
Galison's matrix of beliefs with a (simplified) 
{\em probabilistic network} `Bayesian network'. 
For this reason our belief that also Cinderella is a 
gravitational wave from a BBH merger increases in the 
light that also the sisters are objects of the same kind.
Note that this corroboration effect acts on the priors,
while the Bayes factor should only contain the experimental
information. But this is not exactly true, due to 
role that the priors on the model parameters play in the 
calculation of the Bayes factor via the integral 
$(F.1)$ of footnote \ref{fn:nota_BF}. As soon as we start
getting information about the BBH merger parameters the prior
pdf $f(\underline\theta\,|\,H,I)$ to analyze the next events
becomes less `diffuse' than how they initially were, 
thus increasing the 
value of the integral ($\rightarrow$\,``Occam razor'') 
and then the resulting Bayes factor. 
(For a toy model showing the effect of mutually corroborating 
hypotheses see e.g. the Bayesian network described in Appendix J of
\cite{Columbo}.)  

{\bf Note added}: it is interesting to remark how, after six
months from the first announcement, with much emphasis on the 
sigmas to prove its origin (plus Bayes factors), 
the Monster is finally considered
`self evident', or more precisely,  
``strong enough to be apparent,
without using any waveform model,
in the filtered detector strain data''\,\cite{BBHm_basics}. 
So proceeds Science: the `matrix of belief' has been 
clearly extended.
}
I hope than that LVT151012 will be upgraded to 
GW151012 and that in future searches the Bayes factor
will become the principal figure of merit to rank
gravitational wave candidates. 

I finally conclude with some questions asked at the end of
talk on which this paper is based. 
\begin{itemize}
\item {\em Which Bayes factor would characterize the 750\,GeV excess?}\\
      The result depends on the model to explain the 
      excess\footnote{As an example from Particle Physics of 
      model dependent Bayes factors see \cite{Ghosh}.} 
      and an answer came the week after MaxEnt 2016 by 
      Andrew Fowlie\,\cite{750GeV_BF}. For the model considered
      he got a BF {\em around} 10, the exact value being irrelevant:
      a weak indication, but nothing striking to force
      sceptics to change substantially their  
      opinion.\footnote{A side question is how an experimental team 
      can report the Bayes factor, since it 
      depends on the alternative model. 
      Obviously it cannot (one of ``Laplace's teachings''), 
      but they provide Bayes factors using `popular' models, or 
      it could just report the integral which appears in 
      the denominator, and provide informations that allows
      other physicists to evaluate the numerator, depending on the 
      their model.}
\item {\em Could have CDF at Fermilab claimed 
       to have observed the Higgs boson
       if they had done a Bayesian analysis?}\\
      I am quite positive they could have it, also because 
      the prior on the possible values of the Higgs mass
      was not so vague and well matching the 
      value found later, and therefore the Bayes Factor 
      would have been rather high (and the prior probability
      of a possible manifestation of the boson in the 
      final state was high too).
\end{itemize}
{\bf Acknowledgements}\\
This work was partially supported by a grant from Simons Foundation,
which allowed me a stimulating working environment during
 my visit at the Isaac Newton Institute
of Cambridge, UK. The understanding and/or presentation 
of several things 
of this paper has benefitted 
of the interactions with Pia Astone, Ariel Caticha,
 Kyle Cranmer, Walter Del Pozzo, Norman Fenton,
Enrico Franco, Gianluca Gemme, Stefano Giagu, 
Massimo Giovannini, Keith Inman,
Gianluca Lamanna, Paola Leaci,
Marco Nardecchia, 
Aleandro Nisati, 
and Cristiano Palomba. I am particularly indebded to Allen Caldwell, 
Alvaro de Rujula and John Skilling for many discussions
on physics, probability, epistemology and sociology of 
scientific communities, as well for valuable comments
on the manuscript, which has also benefitted of an accurate reading 
by Christian Durante and 
Dino Esposito.

\end{document}